\documentclass[journal]{IEEEtran}

\usepackage{array}

\usepackage{cite}
\usepackage{amsmath,amssymb,amsfonts}
\DeclareMathOperator*{\argmax}{arg\,max}

\usepackage{algorithmic}
\usepackage{graphicx}
\usepackage{textcomp}
\usepackage{xcolor}
\usepackage{amsthm}
\usepackage{multicol}

\usepackage{stfloats}
\usepackage{url}
\usepackage{verbatim}
\usepackage{algorithm}
\usepackage{enumitem}
\usepackage[caption=false,font=footnotesize]{subfig}

\begin{document}

\title{Integrated Freeway Traffic Control Using Q-Learning with Adjacent Arterial Traffic Considerations 
\thanks{This work has been supported by the METRANS Transportation Center through the Pacific Southwest Region 9 University Transportation AQ:2 Center (USDOT/Caltrans), and through the National Center for Sustainable Transportation
(USDOT/Caltrans).}}

\author{Tianchen Yuan and Petros A. Ioannou~\IEEEmembership{Fellow,~IEEE}
\thanks{T. Yuan and P. Ioannou are with the Department of Electrical Engineering, University of Southern California, Los Angeles, CA, 90089 USA. {\tt\small E-mail: tianchey@usc.edu, ioannou@usc.edu}}}

\maketitle

\begin{abstract}
Numerous studies have shown the effectiveness of intelligent transportation system techniques such as variable speed limit (VSL), lane change (LC) control, and ramp metering (RM) in freeway traffic flow control. The integration of these techniques has the potential to further enhance the traffic operation efficiency of both freeway and adjacent arterial networks. In this regard, we propose a freeway traffic control (FTC) strategy that coordinates VSL, LC, RM actions using a Q-learning (QL) framework which takes into account arterial traffic characteristics. The signal timing and demands of adjacent arterial intersections are incorporated as state variables of the FTC agent. The FTC agent is initially trained offline using a single-section road network, and subsequently deployed online in a connected freeway and arterial simulation network for continuous learning. The arterial network is assumed to be regulated by a traffic-responsive signal control strategy based on a cycle length model. Microscopic simulations demonstrate that the fully-trained FTC agent provides significant reductions in freeway travel time and the number of stops in scenarios with traffic congestion. It clearly outperforms an uncoordinated FTC and a decentralized feedback control strategy. Even though the FTC agent does not control the arterial traffic signals, it leads to shorter average queue lengths at arterial intersections by taking into account the arterial traffic conditions in controlling freeway traffic. These results motivate a future research where the QL framework will also include the control of arterial traffic signals. 

\end{abstract}

\begin{IEEEkeywords}
integrated traffic control, Q-learning, variable speed limit, lane change, ramp metering, traffic signal control.
\end{IEEEkeywords}

\section{Introduction}

\IEEEPARstart{D}{emand} for freeway and arterial travel grows in a fast pace as the population increases in metropolitan areas worldwide, leading to traffic congestion and delays at sensitive parts of road networks such as ramps and intersections. Research efforts on freeway \cite{carlson2010optimal,zhang2018integrated,frejo2020logic} and arterial traffic management \cite{de2016speed,wang2020joint,wang2022coordinated} as separate entities have both achieved certain levels of success in terms of reducing travel time, collision risks and emission rates. However, the integration of freeway and arterial traffic control has been rarely explored due to the difficulty of modeling two completely different traffic patterns and high complexity of the road network. Traditional integration frameworks based on optimization or feedback techniques often face scalability issues or lack of coordination for such tasks \cite{carlson2010optimal,iordanidou2017feedback,zhang2018integrated}. An alternative to these frameworks is reinforcement learning (RL), which has been investigated in some recent studies \cite{schmidt2015decentralised,wang2022coordinated}. The RL framework enhances the coordination by combining the states and actions of sub controllers and offers fast implementation for large-scale road networks. Therefore, in pursuit of integrating freeway and arterial traffic control, we take an initial step to develop a RL-based freeway traffic control strategy that takes into account adjacent arterial traffic conditions.

Popular freeway traffic control techniques include variable speed limit (VSL) control, lane change (LC) control and ramp metering (RM). During the last few decades many research efforts on one or a combination of the above three methods have been proposed to alleviate congestion at freeway bottlenecks. The VSL controller regulates the traffic flow via variable speed commands in order to protect the bottleneck flow to stay at its maximum possible value and reduce the effect of capacity drop \cite{yuan2022selection}. VSL control techniques designed using macroscopic models failed to take into account the forced lane changes at the bottleneck leading to capacity drop which VSL control cannot effectively handle \cite{torne2011evaluation,hadiuzzaman2013cell}. To address this issue, LC recommendations are used at the upstream area of the bottleneck to guide the vehicles onto open lanes in advance and reduce the forced lane changes and the consequential capacity drop \cite{zhang2017combined}. Ramp metering controls the inflow of traffic to the freeway lanes by adjusting the traffic signal timing at each on-ramp entrance. Despite the promising effect of RM in theory \cite{papageorgiou1991alinea,carlson2010optimal}, the on-ramp space is frequently saturated during peak hours, which forces RM to switch off and offers no benefits when it is needed the most. Although alternative solutions which consider the balance of freeway occupancy and on-ramp queue length have been proposed \cite{smaragdis2003series,wang2006local}, the improvement is still limited when both freeway traffic loads and on-ramp demands are high. 

The ramp congestion can be potentially alleviated by designing a freeway traffic control strategy that is adaptive to adjacent arterial traffic conditions. It has been verified primitively in a few studies through the coordination between RM and adjacent arterial signal timing \cite{van2007integrated,haddad2013cooperative,su2014coordinated}. We extend these efforts by integrating VSL, LC, RM actions and taking into account the signal timing and demands of adjacent arterial intersections in the freeway control design using RL algorithms. Considering the unknown transition probability in the traffic environment, model-based RL algorithms is unsolvable. Instead, we adopt model-free RL algorithms such as Q-learning (QL) to search the optimal policy \cite{sutton2018reinforcement}. QL algorithms do not rely on macroscopic models, which tend to be inaccurate in describing traffic behaviors for complex road networks.

The contributions of this paper are summarized as follows:
\begin{enumerate}
    \item We develop an integrated freeway traffic control (FTC) strategy that coordinates VSL, LC, RM actions and considers adjacent arterial traffic conditions using a QL framework. The designed reward function encourages the FTC agent to select control actions that reduce the travel time, the on-ramp queue and stabilize the vehicle density.
    \item We propose a traffic-responsive signal control strategy that computes the arterial signal plan via a simulation-based cycle length model and estimated demands of all intersection approaches. 
    \item The proposed FTC strategy achieves larger benefit in freeway travel time, number of stops and density control compared with uncoordinated control strategies in congested scenarios. The coordination is less effective when the network is not congested. 
    \item Although the proposed FTC strategy does not exercise any control over the arterial traffic signals, it leads to a reduced average queue length at arterial intersections according to the simulation results. The explanation is that the proposed FTC provides a better processing of  off-ramp demands, which interact with arterial traffic, leading to the observed benefits.
\end{enumerate}

The rest of the paper is organized as follows: section \ref{section:litrev} reviews relevant literature. Section \ref{section:method} presents the integrated freeway traffic control strategy and the traffic-responsive arterial signal control strategy. Section \ref{section:Simu} verifies the effectiveness of the proposed approach via microscopic simulations. Section \ref{section:Conclu} concludes the paper.

\section{Literature Review} \label{section:litrev}
The proposed freeway traffic control consists of variable speed limit (VSL) control, lane change (LC) recommendations and ramp metering (RM), while the traffic signal control (TSC) is considered for arterial traffic regulation. We first review relevant research efforts for each control component separately, and then discuss the frameworks to coordinate these components to achieve higher performance. 

Early VSL control designs aimed at stabilizing mainstream traffic flows and minimizing speed variations with reactive rule-based logic \cite{smulders1990control,zackor1991speed}. The reactive nature of these approaches introduces time lags between VSL actions and traffic conditions, and thus, leads to limited performance in terms of travel time and energy consumption. In contrast, many recent developed VSL algorithms compute the speed commands by solving an optimization problem at each time step based on predictions of future traffic states using model predictive control (MPC) techniques \cite{hegyi2003mpc,frejo2014hybrid,khondaker2015variable,muralidharan2015computationally}. The objective function to be optimized typically consists of total travel time, safety measurements, emission rates and fuel consumption. Although MPC-based approaches improve the control performance by eliminating time lags of VSL command activation compared to reactive rule-based approaches, they do not guarantee the stability of vehicle densities and require substantial computational efforts when applied to large-scale road networks. Another well-known alternative is to design a feedback law to compute appropriate speed limits using current and past traffic states \cite{carlson2013comparison,iordanidou2014feedback,jin2015control}, which consumes much less computational efforts than MPC-based approaches. In addition, feedback-based VSL controller guarantees the convergence of mainstream traffic flows and densities analytically \cite{zhang2018stability}. However, feedback-based approaches rely on accurate measurements of traffic states to generate effective control actions. A small deviation from the true value on sensitive variables, vehicle densities for example, may produce unsatisfactory closed-loop behaviors \cite{yuan2021evaluation,alasiri2020robust}. Some reinforcement learning (RL)-based VSL control schemes have been proposed recently as a promising approach to guarantee the optimality via trial and error \cite{walraven2016traffic,li2017reinforcement}. Although the learning process of RL agents can be very time-consuming, the field implementation is fast, and thus, applicable for large-scale road networks.

Lane change control is considered as an effective approach to improve lane-drop bottleneck throughput \cite{alasiri2023per}. However, in most integrated control schemes \cite{zhang2017combined,guo2020integrated,yuan2022selection}, the LC control is designed as rudimentary rules and produces less benefits compared with other controllers such as the VSL. Recently, the LC control has been extended to on-ramp merging bottlenecks using much more mature feedback laws \cite{tajdari2020feedback,kim2023lane}. Both studies \cite{tajdari2020feedback,kim2023lane} assume a connected vehicle environment and balance the lane usage by assigning proper lateral flows. 

Since ramps connect freeways with arterial streets, an appropriate ramp metering design should balance the traffic loads of both regions. Some isolated RM algorithms were first proposed in 1990s \cite{stephanedes1994implementation,zhang1997freeway}, including the famous ALINEA \cite{papageorgiou1991alinea}, which takes freeway occupancy as input and computes the metering rate in a local feedback manner. The classic ALINEA does not consider the potential spillback of on-ramp queues under high traffic demands. Therefore, it was modified in \cite{smaragdis2003series} to avoid the overextension of on-ramp queues by including both the mainstream occupancy and the queue length in the feedback loop. Due to the fact that ramp flows are also affected by the mainstream traffic, coordinated RM algorithms that take into account both local and system-wide traffic conditions typically outperform isolated RM algorithms. In \cite{paesani1997system}, Paesani et al. proposed a system-wide adaptive RM algorithm to compute the metering rates based on estimated future traffic states with linear regression. The lack of accurate real time data makes such methods deviate considerably from the theoretical best performance. In \cite{papamichail2010heuristic}, another extension of ALINEA was made by connecting all the on-ramps via a central controller and dynamically distribute the ramp demands. When one on-ramp queue reaches the threshold, the central controller increases the throughput of this particular on-ramp while decreases the throughput of other on-ramps. In \cite{han2020hierarchical}, a similar two-level structure was embedded into the RM algorithm. The upper level controller determines the optimal total inflow using MPC framework, and the lower level controller distributes the computed total inflow to each on-ramp. Although improvement can be observed by coordinating each RM controller within the network, the control performance is still limited when heavy traffic exists in the mainstream, as RM only affects the vehicle density closely downstream of the on-ramp.

Traffic signal control is considered as the most important and effective method to manage arterial traffic. Existing TSC strategies can be divided into two categories: fixed-time TSC and traffic-responsive TSC. Fixed-time TSC switches between predetermined signal programs according to the time of the day, and thus, suitable for stable, unsaturated traffic conditions. In \cite{webster1958traffic,webster1966traffic}, F. V. Webster designed a signal timing model to minimize the travel delay, and developed the basis for modern fixed-time TSC design. Two of the most widely implemented and extended fixed-time TSC strategies are MAXBAND \cite{little1981maxband} and TRANSYT \cite{robertson1969tansyt}. MAXBAND coordinates traffic signals along an arterial with the same cycle length and proper offsets so that vehicles can travel without stopping, which formulates a progression band with a uniform bandwidth to be maximized. Representative extensions of MAXBAND include assigning multiple bandwidths for different road segments \cite{gartner1991multi}, incorporating route guidance \cite{arsava2016arterial}, or speed advisory \cite{de2016speed}. TRANSYT takes historical traffic data of the road network as input, and then determines the optimal signal control with a heuristic "hill climbing" algorithm. The major limitation of fixed-time TSC is that it cannot handle highly saturated traffic conditions or incident scenarios, which prompts the study of real-time traffic-responsive TSC. In \cite{hunt1982scoot}, a traffic-responsive version of TRANSYT - SCOOT, was developed to adjust signal cycles, splits and offsets with newly-measured traffic flows and occupancy. In \cite{mirchandani2005rhodes}, a real-time hierarchical optimized distributed effective system (RHODES) was proposed with two main operation processes. The first process uses sensor data to estimate future traffic flows within the network. The second process selects the optimal phasing time with dynamic programming (DP) and decision trees. Despite their satisfactory performance in numerous field tests, most traffic-responsive TSC systems rely on accurate real-time traffic data and a powerful central machine to solve optimization algorithms whose complexity grows exponentially with the size of the problem leading to costly implementation with restrictions on the size of arterial networks. 

To coordinate different control components for mixed freeway and arterial road networks, the most intuitive method is to formulate an optimization problem so that all the controllers dedicate to the same objective function \cite{van2007integrated,carlson2010optimal,haddad2013cooperative}. In \cite{van2007integrated,haddad2013cooperative}, separate models are developed to characterize the freeway part and the arterial regions, and then MPC-based control schemes are proposed to minimize the total time spent/delay. Both studies have verified the performance improvement by using a centralized control over the mixed road network, but the considered traffic regulation techniques only involve RM and TSC. Moreover, MPC-based algorithms produce tremendous computational cost for large-scale road networks. Scalable alternatives have been proposed based on shock wave theory, feedback control or logic rules \cite{schelling2011specialist,iordanidou2017feedback,frejo2020logic}. A common drawback of these algorithms is the lack of coordination between different control components, which often leads to suboptimal performance. In \cite{schmidt2015decentralised}, an RL framework is adopted to integrate VSL and RM for freeway traffic control, which guarantees the optimality in theory and is efficient enough for large networks. However, the decentralized control design may lead to insufficient coordination between VSL and RM. In \cite{wang2022coordinated}, a centralized freeway control system that coordinates multiple VSL and RM controllers based on deep RL algorithms is proposed. The considered freeway network is of low complexity. The framework can be potentially extended to involve LC control for a smoother on-ramp merging process. 

\section{Methodology} \label{section:method}
\begin{figure}[h]
\centering
\includegraphics[width = 0.48\textwidth]{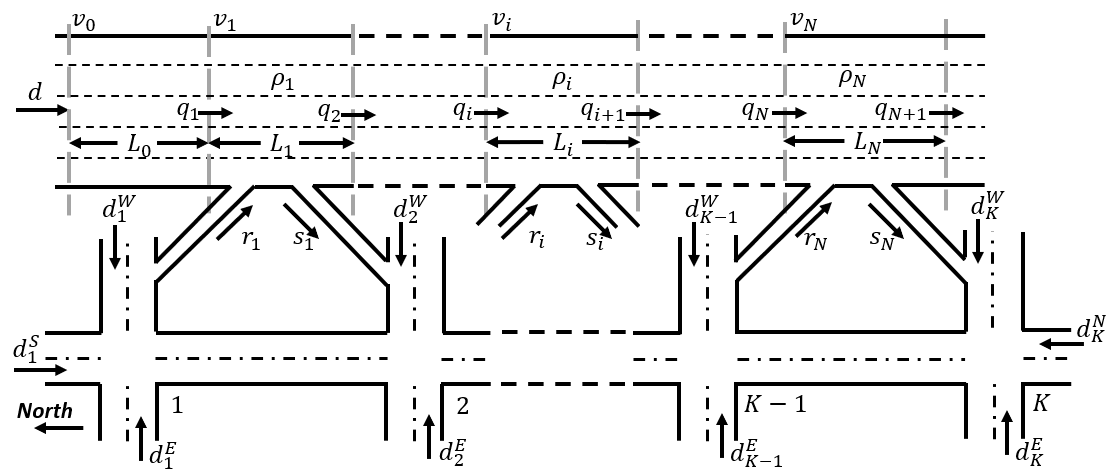}
\caption{Road network consisting of a freeway segment and adjacent arterials.}
\label{fig:roadnet}
\end{figure}

In this section, we propose a freeway traffic control (FTC) strategy that integrates variable speed limit (VSL), lane change (LC), ramp metering (RM) and takes into account adjacent arterial traffic conditions for a connected freeway and arterial road network depicted in Fig. \ref{fig:roadnet}. The purpose of FTC is to reduce average freeway travel time and maintain on-ramp queue lengths and vehicle densities to a reasonable level under all traffic conditions and input demands. A cycle length model modified from the classic Webster model \cite{webster1958traffic,webster1966traffic} is adopted for arterial traffic signal control (TSC). Note that all the freeway ramps are connected with arterial roads. Some connections are omitted in Fig. \ref{fig:roadnet} due to limited drawing space. The freeway segment is divided into $N$ sections where the distance of each section is close to the link distance between two adjacent arterial intersections. More details of the road network configuration will be presented in the numerical simulation. The notations used hereafter are summarized in Table \ref{tab:table_parameters_def}.

\begin{table*}[h]
\centering
\caption{Definition of Variables and Model Parameters}
\label{tab:table_parameters_def}
\begin{tabular}{lll}
\hline
\textbf{Symbol}  & \textbf{Definition} & \textbf{Unit} \\ \hline
$d$              & the freeway vehicle input & veh/h\\
$d^E_k$           & the vehicle input of arterial intersection $k$ with the direction specified by the superscript ($E$ stands for Eastbound) & veh/h\\
$C$              & the capacity of each freeway section & veh/h\\
$C_b$            & the bottleneck capacity      & veh/h    \\
$q_i$           & the mainstream inflow of freeway section $i$     & veh/h \\
$q_{i+1}$       & the mainstream outflow of freeway section $i$     & veh/h \\
$r_i$           & the on-ramp inflow of freeway section $i$     & veh/h \\
$s_i$           & the off-ramp outflow of freeway section $i$     & veh/h \\
$v_i$           & the speed limit command of freeway section $i$        & km/h \\
$v_f$            &  the free flow speed         & km/h    \\
$w$             &   the back propagation speed        & km/h       \\
$\tilde{w}$     &   the rate that the outflow $q_{i+1}$ decreases with density $\rho_i$, when $\rho_i \geq \rho_c$ \cite{srivastava2016lane}  & km/h       \\
$\rho_c$        &     the critical density of the freeway section, at which $v_f\rho_c = w(\rho^j - \rho_c) = \tilde{w}(\tilde{\rho}^j - \rho_c) = C$     & veh/km      \\
$\rho^j$        &   the jam density at which the inflow $q_i$ decreases to 0 with rate $w$ & veh/km\\
$\tilde{\rho}^j$  &   the jam density at which the outflow $q_{i+1}$ decreases to 0 with rate $\tilde{w}$       & veh/km         \\
$\rho_i$        & the density of freeway section $i$        & veh/km \\
$L_i$      &    the length of freeway section $i$       & km   \\
$\epsilon_0$     &    the capacity drop factor, where $\epsilon_0 \in (0,1)$    & unitless     \\ 
\hline
\end{tabular}
\end{table*}

\subsection{Reinforcement Learning} \label{subsec:RL}
The coordination between VSL, LC and RM actions is the key to improving the performance of the proposed integrated freeway control strategy. In typical rule-based or feedback-based control algorithms \cite{iordanidou2017feedback,frejo2020logic,yuan2021evaluation}, each controller attempts to achieve its own goal without coordination. In optimization-based control algorithms \cite{van2007integrated,carlson2010optimal,haddad2013cooperative}, the coordination is addressed as multiple controllers attempting to optimize the same objective. However, the formulation of the optimization problem is tedious for the considered road network due to its high complexity which does not scale up as the network grows in size. 

Reinforcement learning (RL) is a promising alternative approach that enhances the coordination between different controllers and requires less computational effort than optimization-based algorithms in terms of field implementation \cite{schmidt2015decentralised}. The RL agent interacts with the environment and attempts to maximize the cumulative reward by trial and error in discrete time steps, which can be formulated as a Markov decision process (MDP). The MDP contains a set of environment states and a set of control actions. Let $P_a(x,x')$ denote the probability of transition from state $x$ to another state $x'$ by taking action $a$, and let $R_a(x,x')$ denote the reward received from the environment by making the transition from $x$ to $x'$ through action $a$. For each state-action pair $x$ and $a$, the expected discounted reward received by taking action $a$ in state $x$ is expressed as a Q-value function $Q(x,a)$. To solve the MDP problem, we need to find a policy $\pi$ that defines the action that leads to the maximal Q-value in each state, i.e. 
\begin{equation} \label{eq:optpolicyMDP}
    \pi^*(x) = \argmax_a Q^*(x,a)
\end{equation}
where $Q^*(x,a)$ can be expressed as
\begin{equation} \label{eq:q_iter_TP}
    Q^*(x,a) = R_a(x,x') + \gamma\sum_{x'}P_a(x,x') \max_{a'}Q(x',a')
\end{equation}
where $\gamma \in [0,1)$ is a discount factor for the maximum possible future rewards - $\max_{a'}Q(x',a')$. 

The solution of (\ref{eq:optpolicyMDP}) can be calculated using dynamic programming if the transition probability and the reward function are known. However, in the context of traffic flow control, the transition probability cannot be expressed explicitly. Therefore, we apply model-free solution techniques such as the Q-learning (QL) algorithm \cite{watkins1992q} to learn the optimal policy. During the Q-learning process, the Q-value of each state-action pair $(x,a)$ is updated after the agent takes action $a$ at state $x$ and receives an immediate reward $R(x,a)$. Assume $x'$ is the future state for $(x,a)$, equation (\ref{eq:q_iter_TP}) indicates that the optimal Q-value converges to $R_a(x,x') + \gamma \max_{a'}Q(x',a')$. Therefore, $Q(x,a)$ is updated as follows
\begin{equation}
    Q(x, a) \leftarrow Q(x, a) + \eta[R(x, a) + \gamma \max_{a'} Q(x', a') - Q(x, a)]
\end{equation}
where $\eta \in (0,1]$ is the learning rate.

\subsection{Freeway Traffic Control Agent} \label{subsec:ftc}
\begin{figure}[h]
\centering
\includegraphics[width = 0.48\textwidth]{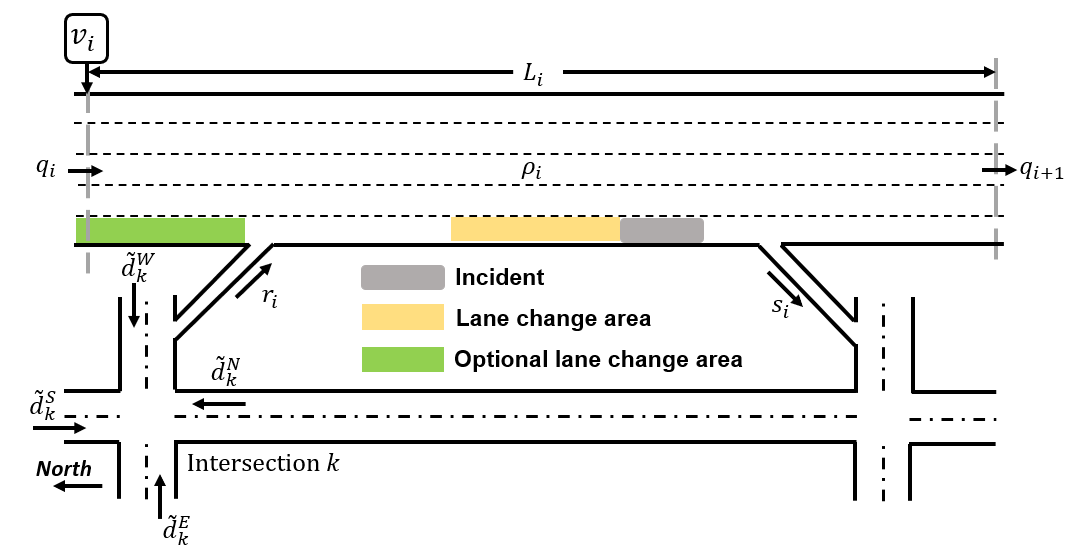}
\caption{Road network of a single freeway section with adjacent arterials.}
\label{fig:roadnetsingle}
\end{figure}

\begin{figure}[h]
\centering
\includegraphics[width = 0.48\textwidth]{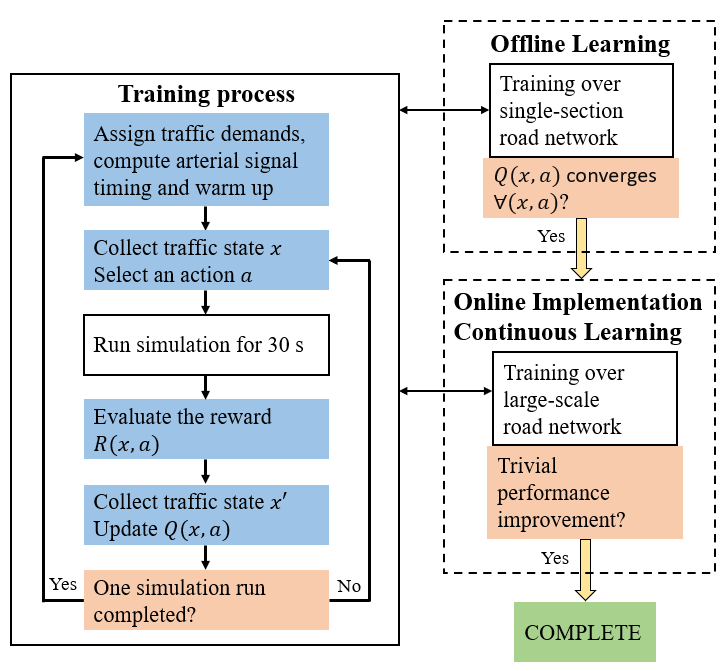}
\caption{The learning process of the freeway traffic control (FTC) agent.}
\label{fig:flowchartQL}
\end{figure}

Based on the QL algorithm presented above, this section aims to design a freeway traffic control (FTC) agent that integrates variable speed limit (VSL), lane change (LC) and ramp metering (RM) actions to smooth on-ramp merging, reduce freeway travel time, and alleviate the congestion produced by a potentially existing lane-drop bottleneck. The FTC agent takes actions based on observations of traffic states within a single-freeway-section environment depicted in Fig. \ref{fig:roadnetsingle}. The adjacent arterial traffic conditions and signal plan are also taken into account to enhance the coordination between the freeway and arterials. The learning process of the FTC agent is presented in Fig. \ref{fig:flowchartQL}. 

As illustrated in Fig. \ref{fig:flowchartQL}, we first perform offline learning over the small road network depicted in Fig. \ref{fig:roadnetsingle}. During the training process, we explore as many state-action pairs as possible by running multiple simulations for each traffic demand level. Each simulation run lasts for 30 min (60 time steps) with a warm-up of 5 min. After the warm-up, there is a 20\% chance that an incident takes place and block the side lane. We consider the offline learning as completed if the Q-value has converged for each state-action pair, which requires the difference between the updated and previous Q-value to be less than 0.01. After that, we implement the trained FTC agent in a large simulation road network with real-world traffic demands. We collect data from the online implementation to assist the continuous learning of the FTC agent. The cycle of online implementation and continuous learning are repeated multiple times until the performance improvement becomes trivial, which means the reduction in travel time and queue lengths between 3 consecutive iterations is less than 1\%. The detailed configurations of the large-scale simulation network is demonstrated in section \ref{subsec:simnetandpara}. 

The states, actions, reward function of the FTC agent and some other important variables and parameters are introduced in the rest of the section.
\subsubsection{State Description}
The states of the FTC agent involve measured and estimated traffic data of the freeway section $i$ and the adjacent arterial intersection $k$, illustrated graphically in Fig. \ref{fig:roadnetsingle} and listed as follows
\begin{itemize}
    \item Vehicle density: $\rho_i$.
    \item Net flow: $q_{i,net}=q_i - q_{i+1} + r_i - s_i$.
    \item On-ramp queue length: $w_i^o$.
    \item The number of closed lane(s): $n_c \in \{0,1\}$.
    \item The dominant signal phase of intersection $k$ in the next control cycle: $n_p \in \{1,2,3,4,5\}$.
    \item Incoming demand of each approach of intersection $k$ estimated by (\ref{eq:estdemand}): $\tilde{d}^E_k$, $\tilde{d}^S_k$, $\tilde{d}^W_k$, $\tilde{d}^N_k$, where the superscript denotes the direction of the approach. The estimation process will be introduced in section \ref{subsec:atm}.
\end{itemize}

To reduce problem dimension, we discretize the state spaces for continuous state variables as follows
\begin{equation} \label{eq:state_discre_FTC}
    \begin{aligned}
        &\rho_i \in \{20,30,40,...,150\} &&\text{veh/km},  \\
        &q_{i,net}, \tilde{d}^E_k, \tilde{d}^S_k, \tilde{d}^W_k, \tilde{d}^N_k \in \{0,100,200,...,4000\} &&\text{veh/h}, \\
        &w_i^o \in \{0,50,100,...,500\} &&\text{m}
    \end{aligned}
\end{equation}

\subsubsection{Action Space}
The action space of the variable speed limit control $A(v_i)$ contains a set of speed limit values that can be applied to each freeway section. Considering the feasibility in the real world, we set the speed limit to be a multiple of 10 km/h with a minimum value of 60 km/h and a maximum value of 100 km/h. The speed limit value can be increased or decreased by at most 10 km/h to ensure safety. Therefore, 
\begin{equation}
\begin{split}
    A(v_i(t)) &= \{\max\{60,v_i(t-T)-10\},\\ &v_i(t-T),\min\{100,v_i(t-T)+10\}\}
\end{split}
\end{equation}
where $T$ is the control cycle.

The ramp metering control has a fixed green phase duration of 3 s and adjusts the red phase duration according to the traffic states. The set of available red phase durations is $\{0,0.5,1,1.5,2,3,4,6\}$ in seconds, which corresponds to the following set of on-ramp flow rates $\{1800,1029,900,800,720,600,514,400\}$ in vehicles per hour. 

To mitigate the capacity drop triggered by forced lane change maneuvers and increase the throughput at the bottleneck, we provide lane change (LC) recommendations to vehicles moving in the closed lane(s) before approaching the bottleneck, as depicted in Fig. \ref{fig:roadnetsingle}. We also provide LC recommendations on the upstream area of on-ramp merging if it improves the overall performance, for which the LC agent needs to make a decision. Therefore, the action space of the LC control is binary, i.e. whether or not to activate the LC recommendations on the upstream area of on-ramp merging. 

\subsubsection{Reward Function}
The objective of the proposed freeway control strategy is to reduce the travel time while maintaining on-ramp queues at reasonable levels and vehicle densities close to the desired value. The average travel time can be computed as \cite{zhang2017combined}
\begin{equation}
    T_t = \frac{1}{N_v} \sum_{j=1}^{N_v} (t_{j,out} - t_{j,in})
\end{equation}
where $N_v$ is the number of vehicles passing through the freeway section during the current control cycle, $t_{j,in}$ and $t_{j,out}$ is the time vehicle $j$ enters and exits the section respectively. 

The above-mentioned control objective requires the reward function to be negatively correlated with the average travel time $T_t$ and the on-ramp queue length $w_i^o$. To avoid on-ramp queue overspill, we want the reward $R = 0$ when $w_i^o$ exceeds the reference value $w_i^r$. Moreover, we want the vehicle density $\rho_i$ to be close to the desired density $\rho^*$, and thus, $R$ should be negatively correlated with the distance between $\rho_i$ and $\rho^*$. Considering the above requirements, the reward function $R$ is defined as
\begin{equation} \label{eq:QLreward}
    R = \max \{0, (1-\frac{w_i}{w_i^r})\frac{L_i}{T_t v_f} - (\frac{\rho_i}{\rho^*}-1)^2 \}
\end{equation}

The reward $R$ defined in (\ref{eq:QLreward}) ranges from 0 to 1. $R$ reaches 0 when the on-ramp queue exceeds the reference value or the density is significantly deviated from the desired value. $R=1$ represents the ideal condition where the freeway section is in free-flow status with the density equal to the desired value and no on-ramp queue. In general, a higher reward value reflects less on-ramp queue length, travel time and deviation between $\rho_i$ and $\rho^*$. Note that the objective of Q-learning is not only to maximize the immediate reward defined in (\ref{eq:QLreward}), but to maximize a cumulative long-term reward where the reward of each time step is computed by (\ref{eq:QLreward}).

\subsubsection{Other Variables and Parameters} \label{subsub:vsl}
As mentioned previously, the reward function (\ref{eq:QLreward}) encourages the density of each freeway section to converge to a predefined value, denoted as $\rho^*$. In the ideal case, a trivial choice is to let $\rho^* = \min \{d,C_b\} / v_f$, which corresponds to the highest possible flow-rate through the bottleneck. However, a small disturbance may drive the density towards the capacity-drop region, which introduces unwanted oscillatory behavior of the closed-loop system and negatively impacts convergence to desired equilibrium states \cite{alasiri2020robust}. To avoid the capacity drop triggered by the disturbance, we multiply $C_b$ with a factor that is slightly less than $1$, and thus, $\rho^* = \min \{d, 0.95C_b\} / v_f$.

The proposed QL algorithm does not cover the control of the upstream part of the freeway segment, which involves two crucial variables - the value and the location of the most upstream VSL sign, denoted as $v_0$ and $L_0$ in Fig. \ref{fig:roadnet}. According to \cite{yuan2022selection}, the inflow $q_1$ is regulated by $v_0$ so that $q_1$ is less than or equal to the bottleneck throughput. The fundamental diagram (FD) indicates that the maximum flow rate produced by $v_0$ is $\frac{v_0 w \rho^j}{v_0 + w}$, which corresponds to three possible values of throughput:
\begin{itemize}
    \item the original capacity $C$ when there is no bottleneck
    \item the bottleneck capacity without capacity drop $C_b$
    \item the bottleneck capacity with capacity drop $(1-\epsilon_0)C_b$
\end{itemize}
Then we have
\begin{equation} \label{eq:v0_VSLcommand}
    v_0 = 
    \left\{
    \begin{aligned}
        &v_f &&\text{no bottleneck}, \\
        &\frac{w C_b}{w\rho^j-C_b} &&\text{bottleneck without capacity drop}, \\
        &\frac{w (1-\epsilon_0)C_b}{w\rho^j-(1-\epsilon_0)C_b} &&\text{bottleneck with capacity drop}
    \end{aligned}
    \right.
\end{equation}
On the other hand, we set $L_0$ to be slightly greater than the lower bound proposed in \cite{yuan2022selection} to reduce backpropagations from the bottleneck if it exists, i.e.
\begin{equation} \label{eq:l0_bound}
    L_0 > \frac{v_f\bar{\rho}(t_0) - (1-\epsilon_0)C_b)v_0 L_b}{((1-\epsilon_0)C_b - v_0\rho_0(t_0))v_f}
\end{equation}
where $\bar{\rho}$ and $L_b$ are the average vehicle density and the distance from the beginning of section 1 to the lane-drop bottleneck respectively; $t_0$ is the time the incident takes place.

In LC control, the distance of the LC area, denoted as $d_{LC}$, is a crucial control variable that needs to be determined properly. $d_{LC}$ must be longer than the minimum distance required for vehicles to complete LC maneuvers safely, but overextending $d_{LC}$ may lead to the underutilization of the road capacity. In this paper, $d_{LC}$ takes an empirical value of 800 m for both on-ramp merging and lane-closure bottlenecks \cite{zhang2017combined}. 

The learning rate $\eta$ is one of the most important QL parameters. A common strategy is to set a decreasing $\eta$ over the training process to ensure reasonable efficiency and the convergence of the Q value. Note that letting $\eta$ decrease as a function of time does not work well because different states and actions are visited at different stages of the learning process. Instead, we assign a specific $\eta$ to each state-action pair and reduce its value every time the state-action pair has been visited \cite{li2017reinforcement}. Therefore,
\begin{equation} \label{eq:Qlearningrate}
    \eta(x,a) = \biggr[\frac{1}{1+n(x,a)(1-\gamma)}\biggr]^{0.8}
\end{equation}
where $n(x,a)$ is the number of times the state-action pair $(x,a)$ has been visited and $\gamma$ is the discount factor. 

To balance exploration and exploitation during the learning process, we implement an adaptive greedy policy where a random action is selected with probability $\delta$ and the best-known action is selected with probability $1-\delta$. Similar to the learning rate equation above, $\delta$ is designed as a function of the number of prior visits to each state \cite{schmidt2015decentralised}, i.e.
\begin{equation} \label{eq:epsgreedyprob}
    \delta(x) = \max \biggl\{0.05, \frac{1}{1+\frac{1}{4N_a(x)}\sum_{n=1}^{a}n(x)} \biggl\}
\end{equation}
where $N_a(x)$ is the number of available actions at the state $x$, $n(x)$ is the number of prior visits of the state $x$. The adopted greedy policy encourages the agent to take random actions (exploration) when a state has not been visited, and becomes more likely to take the best-known action (exploitation) as the number of visits increases. The probability of exploration eventually converges to a minimum value of $0.05$.

\subsection{Arterial Traffic Signal Control} \label{subsec:atm}
The arterial road network under consideration contains $K$ homogeneous signalized intersections indexed from 1 to K in the freeway traffic flow direction, as depicted in Fig. \ref{fig:roadnet}. The on-ramp entrances and off-ramp exits lie on the East side of each intersection. There are $2(K+1)$ entrances plus $N$ off-ramps that generate traffic flow into the arterial road network. At this stage, we assume traffic signal control (TSC) is the only method to regulate arterial traffic flows. Fixed-time TSC strategies cannot fit various input levels and traffic conditions as mentioned in section \ref{section:litrev}. Therefore, we propose a traffic-responsive scheme to determine the signal plan that minimizes the travel time, the fuel consumption and the emissions for each intersection based on the observation of input demands and turning ratios. 

\subsubsection{Cycle Length Model} \label{subsub:optcyclen}
\begin{figure}[h]
\centering
\includegraphics[width = 0.48\textwidth]{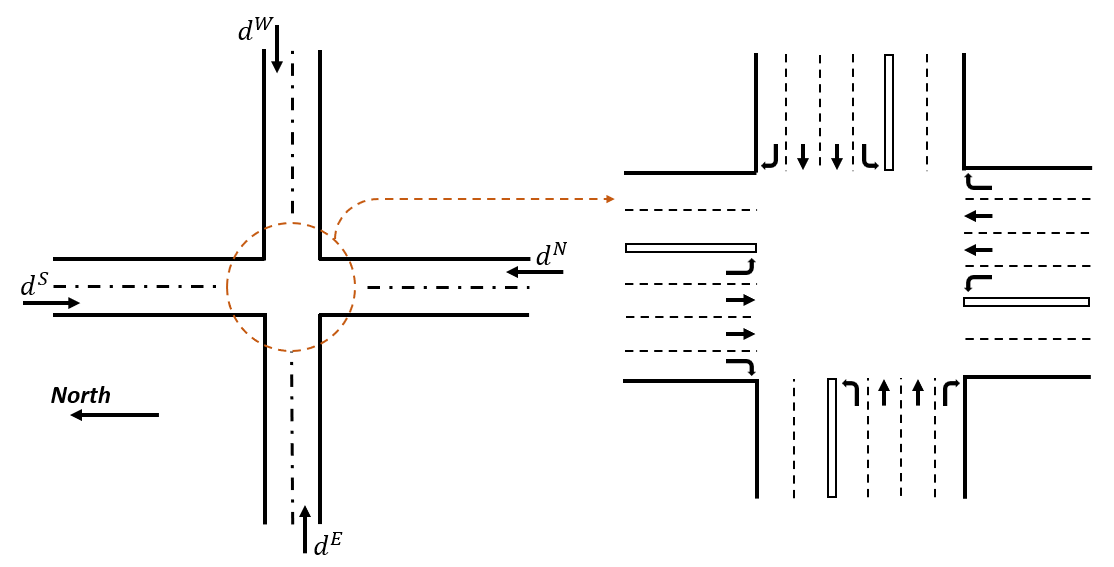}
\caption{The configuration of an isolated intersection.}
\label{fig:intconf}
\end{figure}

\begin{figure}[h]
\centering
\includegraphics[width = 0.48\textwidth]{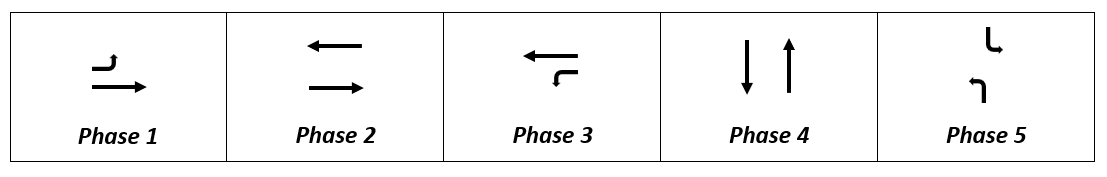}
\caption{The phasing scheme of the default signal plan.}
\label{fig:sigphase}
\end{figure}

The first step of the proposed TSC scheme is to find a cycle length model that fits the arterial traffic conditions. The pioneer research on signal cycle optimization was conducted by F. V. Webster \cite{webster1958traffic,webster1966traffic}, who developed a formula to compute the signal cycle that minimizes travel delays while considering the uncertainties of traffic models as follows:
\begin{equation} \label{eq:webstermodel}
    T_c = \frac{1.5T_l+5}{1-Y}
\end{equation}
where $T_c$ is the signal cycle and $T_l$ is the lost time per cycle. The lost time is defined as the time during which no vehicles are able to pass through an intersection due to the transition between a green phase and a red phase. $Y \in [0,1)$ is the sum of flow ratios of each phase group, which indicates the degree of saturation of an intersection. The flow ratio is defined as the actual traffic flow divided by the saturation flow. The saturation flow is set to be 1800 veh/h/lane in this paper. Extensions based on the Webster model have been made over the years to optimize different objective functions such as fuel consumption, emissions and the number of vehicle stops \cite{li2004signal,hajbabaie2013traffic,calle2019computing}. In this study, we adopt the modified Webster model proposed by Calle-Laguna et al. \cite{calle2019computing}:
\begin{equation} \label{eq:modifiedwebstermodel}
    T_c = \alpha_1 \ln (\frac{T_l}{1-Y}) + \alpha_2
\end{equation}
where $\alpha_1$ and $\alpha_2$ are determined by solving a linear regression problem on the data collected from microscopic simulations over an isolated intersection. The detailed configuration of the isolated intersection is presented in Fig. \ref{fig:intconf}. Each intersection approach is four lanes wide (left, right, and a double through) with a length of 100 m. The arterial road connected to the intersection is two lanes wide and lasts for 1 km on each direction in order to accommodate the potential long queue under high demands. The default signal plan involves five phases as shown in Fig. \ref{fig:sigphase}. Since only medium and high traffic demands are considered, all signal plans must have a separate left-turn phase to enhance the mobility and safety of the intersection operation \cite{bonneson2001evaluating}. 

The commercial microscopic simulator PTV VISSIM 10 is used to calibrate the model parameters $\alpha_1$ and $\alpha_2$ in (\ref{eq:modifiedwebstermodel}). During the calibration, we set the demand space as $d^S,d^E,d^N,d^W \in \{400, 500, 600, ..., 2000\}$ veh/h and the signal cycle space as $T_c \in \{40, 50, 60, ..., 180\}$ s. For each demand level, we evaluate all the cycle options using the following performance index function \cite{li2004signal}:
\begin{equation} \label{eq:optcycperformindex}
    P = \gamma_1 \frac{T_t}{T_{t,0}} + \gamma_2 \frac{F}{F_0} + \gamma_3 \frac{E}{E_0}
\end{equation}
where $T_t$ is the average travel time, $F$ is the average fuel consumption, $E$ is the average emission rates of CO$_2$, and $\gamma_1, \gamma_2, \gamma_3$ are the corresponding weights. $F$ and $E$ are computed using the EPA MOVES model \cite{epa2010motor}. $T_{t,0}, F_0, E_0$ are the base-case results obtained from the scenario where the signal cycle $T_c=60$ s. According to \cite{li2004signal}, we set $\gamma_1=0.4, \gamma_2=\gamma_3=0.3$. 

\begin{figure}[h]
\centering
\includegraphics[width = 0.48\textwidth]{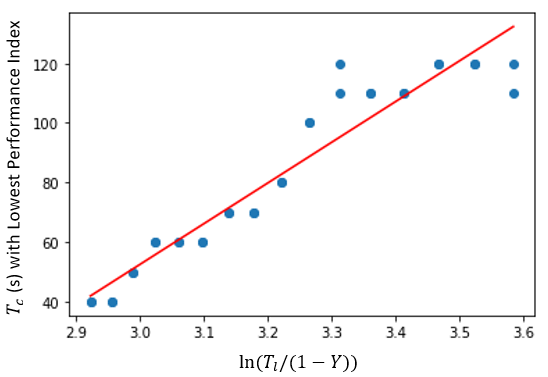}
\caption{Data points and the linear regression curve to determine the cycle length model for the considered intersection.}
\label{fig:linregsigcyc}
\end{figure}

After iterating the data collection process for all demand levels twice, we plot the data points with the best (lowest) performance index and perform the linear regression in Fig. \ref{fig:linregsigcyc}. As a result, $\alpha_1 = 136.8, \alpha_2 = -357.7$.  

\subsubsection{Demand and Flow Ratio Estimation}
To determine signal plans using the obtained cycle length model, we estimate the demands of each intersection as follows:
\begin{equation} \label{eq:estdemand}
    \begin{aligned}
    \tilde{d}_k^W &= d_k^W + \bar{s}_i\\
    \tilde{d}_k^E &= d_k^E\\
    \tilde{d}_k^S &= \left\{ 
        \begin{aligned}
        &d_1^S &&\text{if } k = 1,\\
        &y_{k-1}^{W,l}\tilde{d}_{k-1}^W + y_{k-1}^{E,r}\tilde{d}_{k-1}^E + y_{k-1}^{S,t}\tilde{d}_{k-1}^S && \text{otherwise } \\
        \end{aligned}
        \right. \\
    \tilde{d}_k^N &= \left\{ 
        \begin{aligned}
        &d_K^N &&\text{if } k = K,\\
        &y_{k+1}^{W,r}\tilde{d}_{k+1}^W + y_{k+1}^{E,l}\tilde{d}_{k+1}^E + y_{k+1}^{N,t}\tilde{d}_{k+1}^N && \text{otherwise } \\
        \end{aligned}
        \right.
    \end{aligned}
\end{equation}
where $\tilde{d}_k^W$ is the estimated demand of the Westbound approach at intersection $k$, $d_k^W$ is the actual Westbound vehicle input of intersection $k$, $\bar{s}_i$ is the historical average off-ramp flow rate of freeway section $i$ ($\bar{s}_i=0$ if the off-ramp does not exist), $y_k^{W,l}$ is the left-turn ratio of the Westbound approach at intersection $k$. The superscript denotes the traffic flow direction in the following manner: $E$ - Eastbound, $W$ - Westbound, $N$ - Northbound, $S$ - Southbound, $l$ - left-turn, $r$ - right-turn, $t$ - through. We assume freeway section $i$ is associated with intersection $k$, all vehicle inputs are known and all links within the arterial network are unsaturated. The estimation process of turning ratios will be illustrated in section \ref{subsec:simnetandpara}.

Then we calculate the flow ratio of each phase group with respect to the phase scheme presented in Fig. \ref{fig:sigphase} and sum them up to obtain $Y$:
\begin{equation} \label{eq:calcflowratio}
    \begin{aligned}
    Y_1 &= \frac{\tilde{d}^S}{q_s^S} \\
    Y_2 &= \frac{(y^{S,r}+y^{S,t})\tilde{d}^S}{q_s^S} + \frac{(y^{N,r}+y^{N,t})\tilde{d}^N}{q_s^N} \\
    Y_3 &= \frac{\tilde{d}^N}{q_s^N} \\
    Y_4 &= \frac{(y^{W,r}+y^{W,t})\tilde{d}^W}{q_s^W} + \frac{(y^{E,r}+y^{E,t})\tilde{d}^E}{q_s^E} \\
    Y_5 &= \frac{y^{W,l}\tilde{d}^W}{q_s^W} + \frac{y^{E,l}\tilde{d}^E}{q_s^E} \\
    Y &= Y_1 + Y_2 + Y_3 + Y_4 +Y_5
    \end{aligned}
\end{equation}
where $q_s$ is the saturation flow of the approach whose direction is specified by the superscript. In this paper, $q_s = 7200$ veh/h for all directions at each intersection. Since (\ref{eq:calcflowratio}) applies to all intersections in the road network, the index of the intersection is omitted for the sake of simplicity. 

\subsubsection{Cycle Length and Split Computation}
We compute the signal cycle for each intersection using (\ref{eq:modifiedwebstermodel}) and then select the closest value as the actual cycle $T_c$ from the cycle space $\{40, 50, 60, ..., 180\}$ s. If the result of (\ref{eq:modifiedwebstermodel}) happens to be in the middle of two options, we select the larger one.

Once $T_c$ is determined, we can allocate the green light time for each phase according to the flow ratios found in (\ref{eq:calcflowratio}):
\begin{equation}
    T_{g,j} = \frac{(T_c - T_l)Y_j}{Y}, \quad \text{for} \ j = 1,2,3,4,5,
\end{equation}
where $T_{g,j}$ denotes the green light time of phase $j$ per cycle, and the lost time $T_l = 16$ s. 

To minimize travel delays and improve the traffic mobility, it is recommended to unify the cycle length for closely spaced traffic signal and use proper offsets to create a progression band (green wave) for vehicle platoons on the main street \cite{bonneson2001evaluating}. However, the intersections in our simulation network are relatively far apart with an average distance of 1.5 km. Besides, the longitudinal traffic is not significantly larger than the lateral traffic at each intersection. Therefore, the offset optimization is not considered in this paper. The offset of each signal is simply set to 0 s.

\section{Numerical Simulations} \label{section:Simu}

\subsection{Simulation Network and Parameters} \label{subsec:simnetandpara}
\begin{figure*}[h]
\centering
\includegraphics[width=0.96\linewidth]{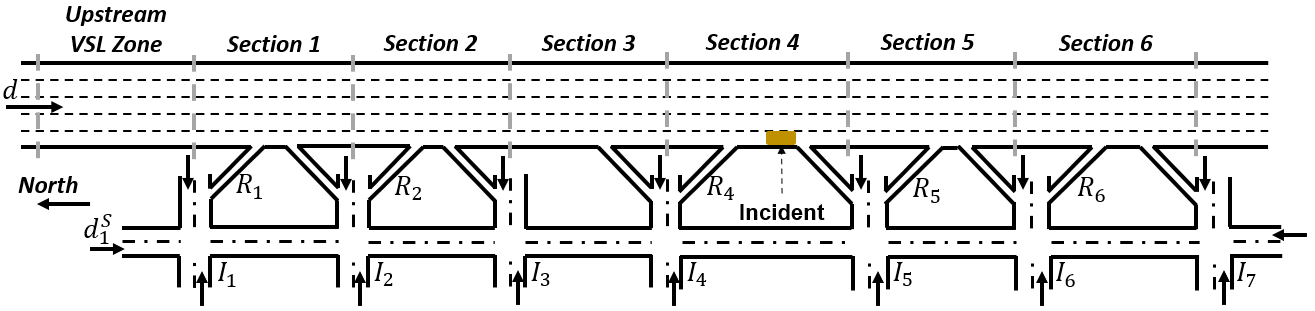}
\caption{The simulation road network of I-710 freeway and the adjacent arterials. The gold rectangle denotes the location of the incident in some scenarios.}
\label{fig:simnet}
\end{figure*}

\begin{figure*}[h]
\centering
\includegraphics[width=0.96\linewidth]{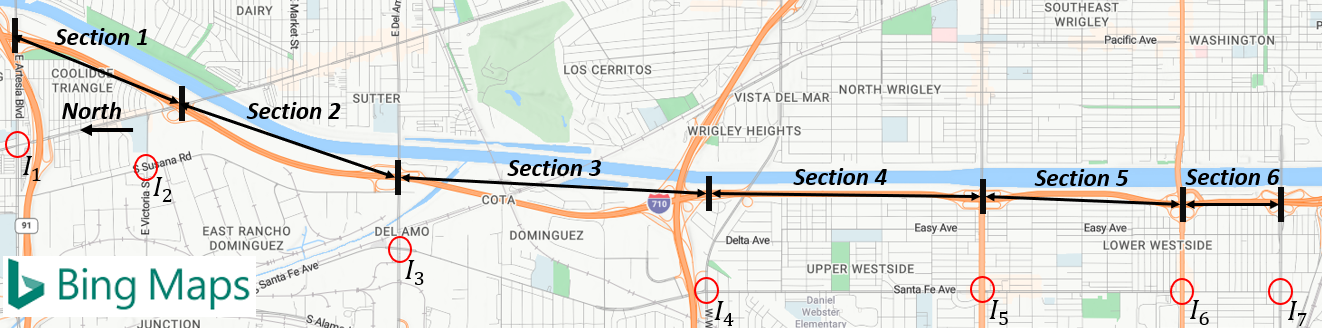}
\caption{The simulation road network of I-710 freeway and the adjacent arterials on Bing Map. The upstream VSL zone is omitted to save the space.}
\label{fig:simnetBing}
\end{figure*}

The proposed control methodologies are simulated using a microscopic traffic simulator based on the commercial software PTV Vissim 10. The road network in Fig. \ref{fig:simnet} contains a 16-km segment of I-710 freeway and the adjacent arterial region in Los Angeles, California, United States. The freeway segment is divided into 6 sections and one upstream VSL zone. The length of each freeway section is slightly longer than the distance of the parallel arterial link due to its curved shape, with an average of 1.6 km. The length of the upstream VSL zone is 4 km and the deployment location of $v_0$ is determined by (\ref{eq:l0_bound}). The freeway segment has 5 lanes, 5 on-ramps and 6 off-ramps. All ramps are connected with the arterial road network. There are 7 arterial intersections aligned in parallel with the freeway. The real-world location of each intersection is marked by a red circle in Fig. \ref{fig:simnetBing}. The selected intersections are all major intersections whose traffic conditions are closely related to the states of adjacent freeway section. We simplify the simulation network by directly linking some of the intersections and ignoring vertical freeways. 

Traffic demands are generated from one freeway entrance and 16 arterial entrances as indicated by arrows in Fig. \ref{fig:simnet}. The freeway demand is based on hourly average traffic volume data of April 2019 from the Caltrans Performance Measurement System (PeMS). The arterial demands are based on hourly traffic counts data of April 2019 from the LADOT Database. We consider two levels of traffic demands. The moderate level is based on the average hourly traffic data of the whole month. The high-level demand is 40\% more than the moderate-level demand. Each simulation run lasts for 40 min. If the incident exists, it takes place after a 10-min warm-up and will be cleared at 30 min.

Based on our observations, we've noted that the saturation demand is approximately double the moderate-level demand. This heightened demand often leads to long queues at intersections, occasionally causing spillback onto the freeway via off-ramps. Our proposed approach does not effectively address this situation, highlighting the need for further research.

The turning ratio is determined by the ratio of the traffic flow of each direction, where each traffic flow follows a normal distribution $\mathcal{N}(\mu, \sigma^2)$. $\mu$ and $\sigma$ are the mean and the standard deviation of the LADOT traffic data of the corresponding direction. Although the actual turning ratio is unknown, we use the mean value of the flow distribution to estimate the turning ratio of an intersection approach as follows:
\begin{equation}
    y^t = \frac{\mu^t}{\mu^l + \mu^t + \mu^r}
\end{equation}
where $l,t,r$ stand for left-turn, through, right-turn respectively. The turning ratios of on-ramps and off-ramps are estimated in the same manner as those of approaches.  

The above-mentioned road network and parameter setting are applied to the online implementation during the learning process of the freeway traffic control (FTC) agent. There are 4 scenarios to be executed in one online implementation - moderate demands without incident, moderate demands with incident, high demands without incident, high demands with incident. The cycle of online implementation and continuous learning is iterated until the performance improvement in travel time and queue lengths between 3 consecutive iterations is less than 1\%. Then we apply the refined FTC agent for each scenario once more to obtain the final simulation results presented in section \ref{subsec:simres}. In addition, we are interested in 3 other types of freeway control as comparisons. The arterial traffic signal control (TSC) is given and is known to the FTC agent. All freeway control strategies to be tested are summarized as follows:
\begin{enumerate}[label=(\roman*)]
    \item No freeway control: inactive VSL, LC and RM control.
    \item Decentralized feedback control: a decentralized feedback control strategy proposed in \cite{zhang2018integrated}.
    \item QL without coordination: uncoordinated FTC trained by a similar QL algorithm. To eliminate the coordination, the FTC agent is divided into three sub agents (VSL, LC, RM) for separate training. Each agent has only one action variable (e.g. $v_i$ for VSL). In addition, none of these agents considers the adjacent arterial traffic conditions, and thus, $n_p, \tilde{d}^E_k$, $\tilde{d}^S_k$, $\tilde{d}^W_k$, $\tilde{d}^N_k$ are removed from the state variables.
    \item QL with coordination: the proposed FTC.
\end{enumerate}

\subsection{Simulation Results} \label{subsec:simres}
In this section, we evaluate the performance of 4 types of control strategies as previously mentioned in the road network depicted in Fig. \ref{fig:simnet} for each interested scenario. The performance criteria are listed as follows \cite{zhang2017combined}: 
\begin{itemize} 
    \item Freeway average travel time ($T_t$): the average time spent for each vehicle to travel through the freeway segment.
    \begin{equation}
        T_t = \frac{1}{N_v}\sum_{i=1}^{N_v} (t_{i,out}-t_{i,in}) 
    \end{equation}
    where $N_v$ is the number of vehicles passing through the freeway segment, $t_{i,in}$ and $t_{i,out}$ is the time vehicle $i$ enters and exits the freeway segment respectively. $t_{i,in} > 10$ min so that the warm-up period is excluded. Vehicles that enter or exit from ramps are also excluded. 
    \item Freeway average number of stops ($\bar{s}$): the average number of stops performed by each vehicle when traveling through the freeway segment.
    \begin{equation}
        \bar{s} = \frac{1}{N_v}\sum_{i=1}^{N_v} s_i
    \end{equation}
    where $s_i$ is the number of stops performed by vehicle $i$. The warm-up period is excluded.
    \item Freeway average emission rates of CO$_2$ ($E$): the calculation of emission rates is based on the MOVES model provided by the Environment Protection Agency \cite{epa2010motor}. 
    \begin{equation}
        E = \sum_{i=1}^{N_v} E_i / \sum_{i=1}^{N_v} l_i
    \end{equation}
    where $E_i$ is the emission produced by vehicle $i$ and $l_i$ is the travelled distance of vehicle $i$. The warm-up period is excluded.
    \item Average on-ramp queue length ($\bar{w}_o$):
    \begin{equation}
        \bar{w}_o = \sum_{i=1}^{N} \bar{w}_i^o / N_o
    \end{equation}
    where $N$ is the number of freeway sections, $\bar{w}_i^o$ is the average queue length of on-ramp $i$ during the simulation except the warm-up, $N_o$ is the number of on-ramps.
    \item Average queue length of arterial intersections ($\bar{w}_a$):
    \begin{equation}
        \bar{w}_a = \sum_{k=1}^{K} (\bar{w}^N_k + \bar{w}^S_k + \bar{w}^E_k + \bar{w}^W_k) / 4K
    \end{equation}
    where $K$ is the number of arterial intersections, $\bar{w}^N_k$ is the average queue length of the Northbound approach of intersection $k$ during the simulation except the warm-up.
\end{itemize}

Considering the stochastic nature of microscopic simulations, we take the average of 10 simulation runs for each pair of control type and scenario and record the final results in Table \ref{tb:eval_mowo}, \ref{tb:eval_mowi}, \ref{tb:eval_hiwo} and \ref{tb:eval_hiwi}. 

\begin{table*}[h]
\begin{center}
\caption{Evaluations of a Moderate-Demand Scenario without Incident} \label{tb:eval_mowo}
\begin{tabular}{cccccc}
Control & $T_t$ (min) & $\bar{s}$ & $E$ (g/veh/km) & $\bar{w}_o$ (m) & $\bar{w}_a (m)$    \\ \hline
No freeway control & 10.2 & 0.3 & 200.5 & 0 & 11.2                 \\
Decentralized feedback control & 10.2 & 0.2 & 199.8 & 0 & 11.9                  \\
QL without coordination & 10.2 & 0.2 & 199.4 & 0 & 12.5                      \\
QL with coordination & 10.2 & 0.2 & 200 & 0 & 11.6                   \\
\hline
\end{tabular}
\end{center}
\end{table*}

\begin{table*}[h]
\begin{center}
\caption{Evaluations of a Moderate-Demand Scenario with Incident} \label{tb:eval_mowi}
\begin{tabular}{cccccc}
Control & $T_t$ (min) & $\bar{s}$ & $E$ (g/veh/km) & $\bar{w}_o$ (m) & $\bar{w}_a (m)$    \\ \hline
No freeway control & 12.3 & 0.7 & 212.7 & 0 & 14.5                 \\
Decentralized feedback control & 11.4(7\%) & 0.5(29\%) & 209(2\%) & 13.1 & 11.7                  \\
QL without coordination & 11.3(8\%) & 0.5(29\%) & 208.1(2\%) & 15.9 & 11.6                      \\
QL with coordination & 10.8(12\%) & 0.3(57\%) & 203.2(4\%) & 16.4 & 11.7                   \\
\hline
\end{tabular}
\end{center}
\end{table*}

\begin{table*}[h]
\begin{center}
\caption{Evaluations of a High-Demand Scenario without Incident} \label{tb:eval_hiwo}
\begin{tabular}{cccccc}
Control & $T_t$ (min) & $\bar{s}$ & $E$ (g/veh/km) & $\bar{w}_o$ (m) & $\bar{w}_a (m)$    \\ \hline
No freeway control&12&2.1&241.7&25.3&49.6\\
Decentralized feedback control&11.7(2\%)&1.6(24\%)&233.7(3\%)&30.2&49.3\\
QL without coordination&11.6(3\%)&1.7(19\%)&232.5(4\%)&44.1&47.8\\
QL with coordination&11.3(6\%)&1.4(33\%)&225.5(7\%)&40.8&45.6\\
\hline
\end{tabular}
\end{center}
\end{table*}

\begin{table*}[h]
\begin{center}
\caption{Evaluations of a High-Demand Scenario with Incident} \label{tb:eval_hiwi}
\begin{tabular}{cccccc}
Control & $T_t$ (min) & $\bar{s}$ & $E$ (g/veh/km) & $\bar{w}_o$ (m) & $\bar{w}_a (m)$    \\ \hline
No freeway control&14.1&4.9&262.4&33&56.7\\
Decentralized feedback control&13(8\%)&2.9(41\%)&253(4\%)&35.9&55.2\\
QL without coordination&13.3(6\%)&3.2(35\%)&254.2(3\%)&48.7&51\\
QL with coordination&12.4(12\%)&2.4(51\%)&245.3(7\%)&47&47.5\\
\hline
\end{tabular}
\end{center}
\end{table*}

Table \ref{tb:eval_mowo} presents the results of a moderate-demand scenario without incident, where there is no bottleneck on freeway and minimal control effort is needed. Thus, the results of all types of control are close to each other. The average number of stops is supposed to be 0 in reality. However, the stop happens in simulations when two vehicles are very close to each other at ramp merging points. Table \ref{tb:eval_mowi} presents the results of a moderate-demand scenario with incident, where the incident introduces a lane-drop bottleneck that increases the travel time ($T_t$) and the number of stops ($\bar{s}$) significantly. The percentages in brackets quantify the performance improvement by implementing the corresponding control scheme versus no freeway control. The percentage improvement of the coordinated control is higher than decentralized feedback control and the uncoordinated control, which verifies the benefit of integrating VSL, LC and RM actions. 

Table \ref{tb:eval_hiwo} presents the results of a high-demand scenario without incident, where the increased demand introduces bottlenecks at on-ramp merging areas occasionally. The overall performance improvement by freeway traffic control is less obvious compared with Table \ref{tb:eval_mowi} because the ramp-merging bottleneck is less detrimental than the lane-drop bottleneck. The coordinated control still outperforms uncoordinated control schemes with less margins. Table \ref{tb:eval_hiwi} presents the results of a high-demand scenario with incident, where both the lane-drop bottleneck and ramp-merging bottlenecks exist. The performance margins between the coordinated and uncoordinated control schemes increase as we introduce the incident. The above observations indicate that the proposed coordination mechanism provides higher benefits when the demand grows up or when the incident takes place. 

A trade-off between the freeway performance ($T_t,\bar{s},E$) and on-ramp queue lengths ($\bar{w}_o$) can be observed by using any type of freeway traffic control. Taking the coordinated control in Table \ref{tb:eval_hiwi} as an example, the average on-ramp queue is 14 m longer while the average freeway travel time is reduced by 12\%. The increase in the queue length is trivial considering the minimum on-ramp queue capacity, which is over 300 m. Therefore, the trade-off is acceptable. By considering the adjacent arterial traffic conditions, the coordinated control demonstrates slightly better performance compared to the uncoordinated QL algorithm in managing on-ramp queue lengths. This improvement may stem from an earlier activation of lane change recommendations, which smooths the merging process for large on-ramp demands.

Despite the fact that the FTC agent does not control the arterial traffic signals, the coordinated freeway control leads to a significant reduction in average queue length of arterial intersections ($\bar{w}_a$), possibly due to a better processing of the off-ramp demands feeding into the arterial network. This benefit motivates a fully integrated control approach where the QL framework also involves the control of arterial traffic signals - a problem that is currently under investigation. 

\begin{figure}[h]
    \centering
  \subfloat[\label{fig:moworho4}]{%
       \includegraphics[width=0.48\linewidth]{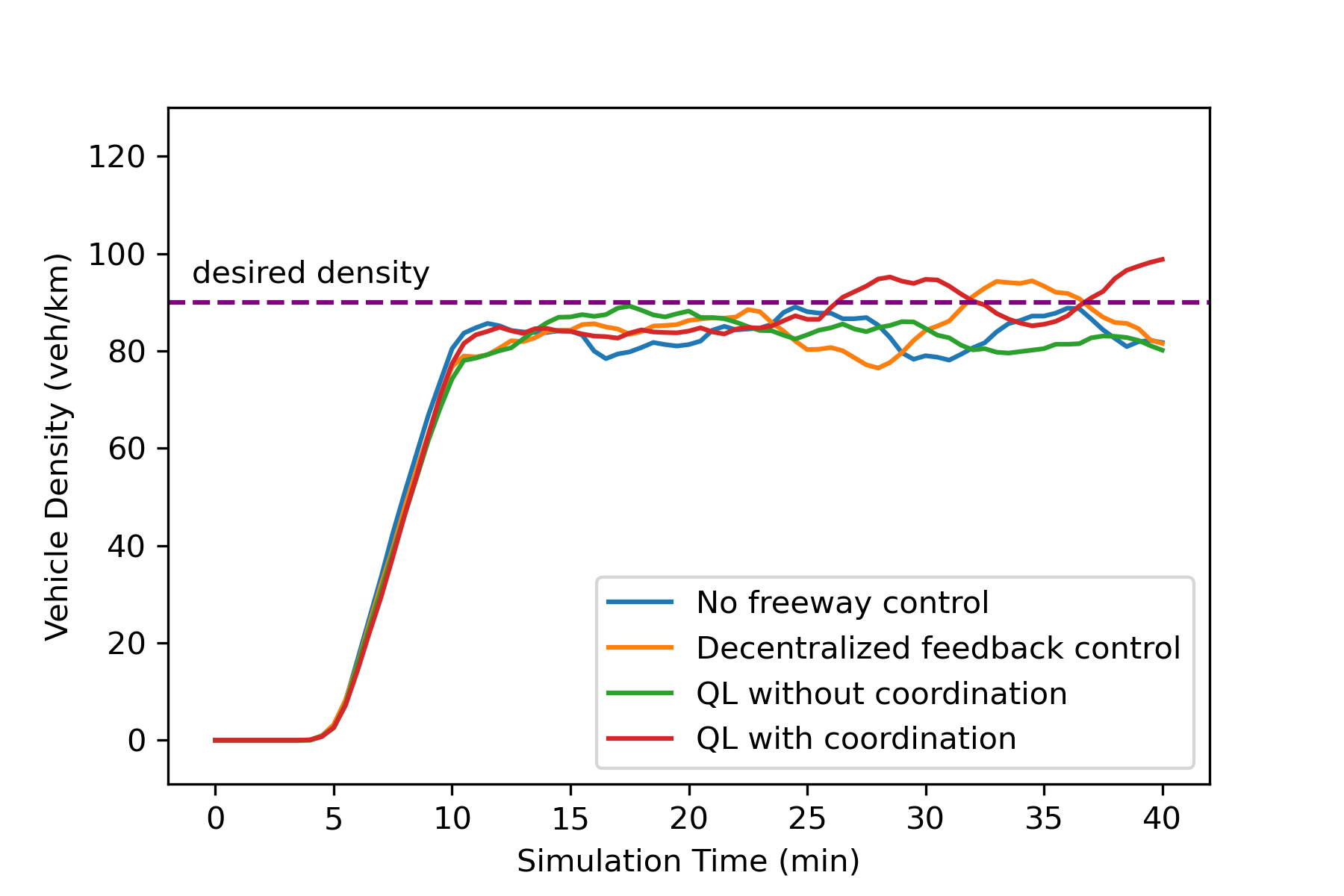}}
    \hfill
  \subfloat[\label{fig:mowirho4}]{%
        \includegraphics[width=0.48\linewidth]{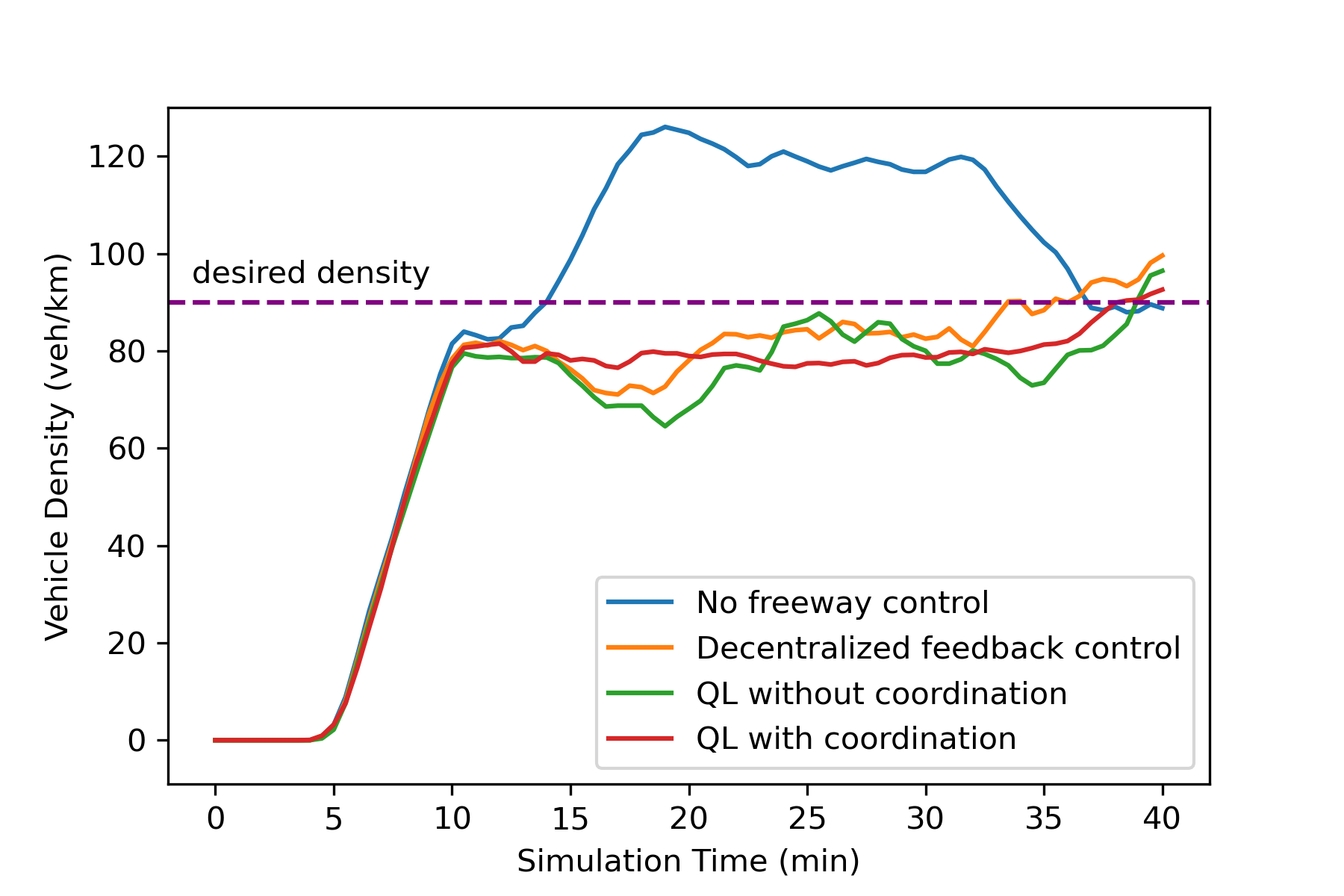}}
    \\
  \subfloat[\label{fig:hiworho4}]{%
        \includegraphics[width=0.48\linewidth]{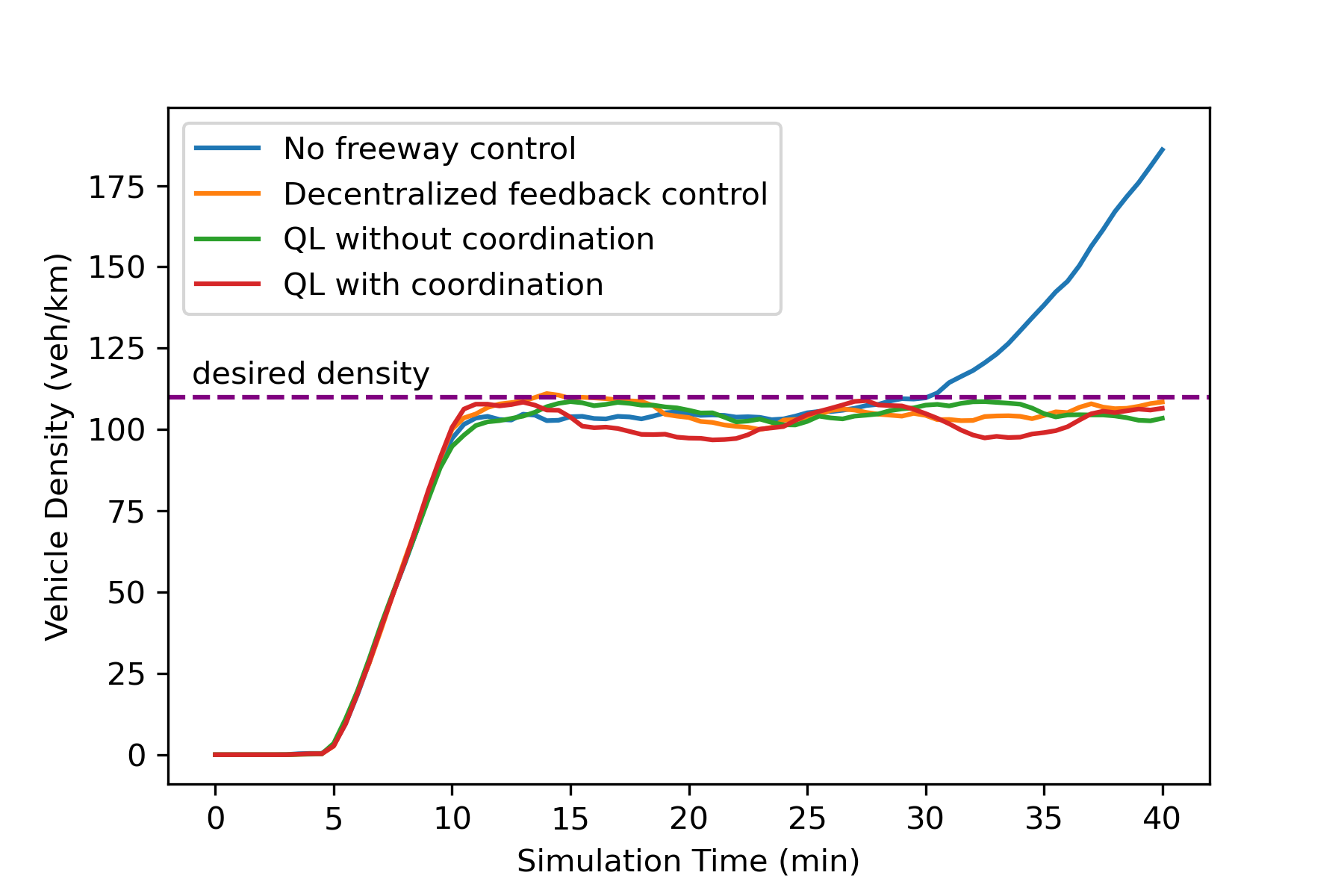}}
    \hfill
  \subfloat[\label{fig:hiwirho4}]{%
        \includegraphics[width=0.48\linewidth]{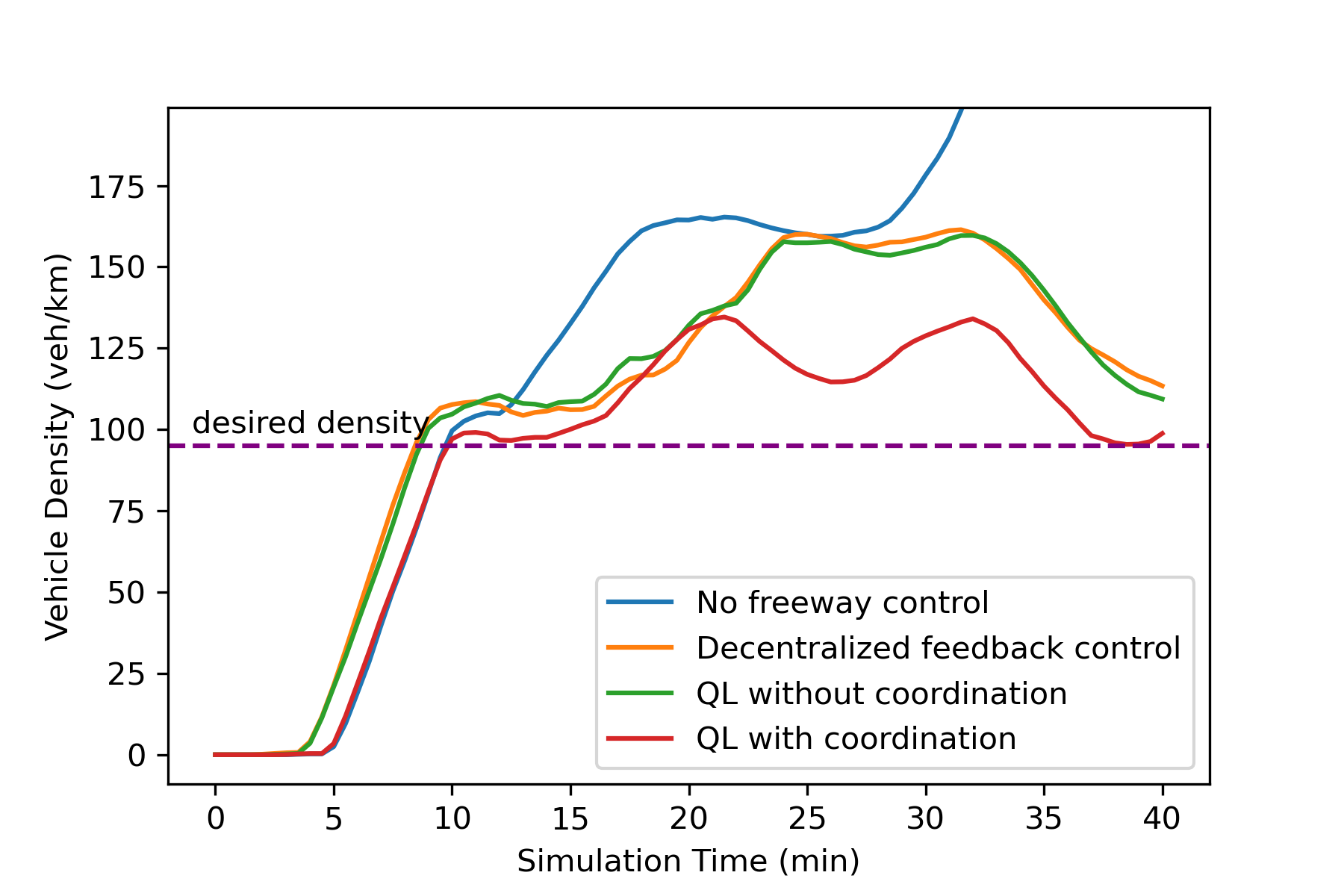}}
  \caption{Measured vehicle density profiles of each scenario in freeway section 4 (where the incident takes place). (a) moderate demand without incident, (b) moderate demand with incident, (c) high demand without incident, (d) high demand with incident.}
  \label{fig:denprofsec4} 
\end{figure}

We draw density profiles for freeway section 4 in Fig. \ref{fig:denprofsec4} as a representative since the density profiles of each section shares similar behaviors. The purple dash line marks the desired density $\rho^*$ in each scenario. We spend a 10-min warm-up loading traffic for the entire network. Then the control algorithm kicks in and maintains the density of each freeway section at the steady state. Fig. \ref{fig:denprofsec4} shows that the interested control schemes deliver similar performance in terms of stabilizing the density when there is no incident. The coordinated control performs significantly better than the uncoordinated control and decentralized feedback control when the incident occurs in high-demand scenarios. The above results also demonstrate the necessity of coordinating different control components when the road network is congested.

\section{Conclusion} \label{section:Conclu}
In this paper, we proposed an integrated freeway traffic control (FTC) strategy that coordinates variable speed limit (VSL), lane change (LC), ramp metering (RM) actions using a Q-learning (QL) framework. The considered road network consists of both a freeway segment and adjacent arterial roads. The arterial signal timing and intersection demands are included as state variables of the FTC agent to improve the control performance. The FTC agent is trained offline in a single-section road network first, and then implemented online in the I-710 simulation road network with real-world traffic demands from PeMS and LADOT database. The online data are collected to assist the continuous learning of the agent. The numerical simulations indicate that the proposed FTC achieves a 3-6\% performance margin in freeway travel time compared to uncoordinated control or decentralized feedback control in high-demand or incident scenarios. Even though the FTC agent does not control arterial traffic signals, it does lead to shorter queues at arterial intersections by taking into account the arterial traffic and the demand generated at freeway on-ramps and off-ramps. The extension of the proposed approach involves incorporating the control of arterial traffic signals within the QL framework, which is currently under investigation.



%\newpage


%\vfill


\end{document}